\documentclass{article}
\usepackage{ae,aecompl}

\usepackage[T1]{fontenc}
\usepackage[latin9]{inputenc}
\usepackage{amsmath}
\usepackage{amsthm}
\usepackage{amssymb}
\usepackage{stmaryrd}

\makeatletter
\numberwithin{equation}{section}
\numberwithin{figure}{section}
\theoremstyle{plain}
\newtheorem{thm}{\protect\theoremname}
  \theoremstyle{plain}
  \newtheorem{lem}[thm]{\protect\lemmaname}
  \newtheorem{prop}[thm]{\protect\propname}
  \theoremstyle{remark}
  \newtheorem{rem}{\protect\remarkname}

\usepackage{amsmath,epsfig,stmaryrd}

\makeatother

  \providecommand{\lemmaname}{Lemma}
  \providecommand{\remarkname}{Remark}
\providecommand{\theoremname}{Theorem}
\providecommand{\propname}{Proposition}
\title{On the non-detectability of spiked large random tensors}

\author{A. Chevreuil and P. Loubaton}

\date{Laboratoire d'Informatique Gaspard Monge (CNRS, Université Paris-Est/MLV)\\ 
	5 Bd. Descartes 77454 Marne-la-Vallée (France)}

\begin{document}

\maketitle

\begin{abstract}
This paper addresses the detection of a low rank  high-dimensional tensor corrupted 
by an additive complex Gaussian noise. In the asymptotic regime where 
all the dimensions of the tensor converge towards $+\infty$ at the same rate, existing 
results devoted to rank 1 tensors are extended. It is proved
that if a certain parameter depending on the low rank tensor is below a threshold, then
the null hypothesis and the presence of the low rank tensor are undistinguishable hypotheses in the sense
that no test performs better than a random choice. 
\end{abstract}

\section{Introduction}

\label{sec:intro}

The problem of testing whether an observed $n_{1}\times n_{2}$ matrix ${\bf Y}$ 
is either a zero-mean independent identically distributed Gaussian
random matrix ${\bf Z}$ with variance $\frac{1}{n_{2}}$, or ${\bf X}_{0} +{\bf Z}$ (where ${\bf X}_{0}$ is a low rank matrix: a
	 useful signal, called also \textit{spike})  is a fundamental problem arising in numerous applications
such as the detection of low-rank multivariate signals or the Gaussian
hidden clique problem. 
When the two dimensions $n_1,n_2$ converge towards $\infty$ at the same rate,  the rank of 
${\bf X}_{0}$ remaining fixed, the context is this of the so-called  additive spiked large random matrix models. 
Various results on the singular values of ${\bf X}_{0} +{\bf Z}$ have been established; in particular it is possible to show
that the Generalized Likelihood Ratio Test (GLRT)
is consistent (i.e. both the probability of false alarm and the probability of missed detection both converge towards $0$ when $n_{1},n_{2}$ converge towards $+\infty$ in such a way that $n_{1}/n_{2}\rightarrow c>0$)
if and only if and only if the largest singular value of $\mathbf{X}_{0}$
is above the threshold $c^{1/4}$ (see e.g.
\cite{nadakuditi-edelman-2008}, 
\cite{bianchi-et-al-2011}, \cite{benaych2012singular}).

In a number of real life problems, the observation is not a matrix,
but a tensor ${\bf Y}$ of order $d\geq3$, i.e. a $d$\textendash dimensional
array ${\bf Y}={\bf Y}_{i_{1},i_{2},\ldots,i_{d}}$ where for each
$k=1,\ldots,d$, $i_{k}\in[1,\ldots,n_{k}]$. In this context, the
generalization of the above matrix hypothesis testing problem becomes:
test that the observed order $d\geq3$ tensor is either a zero-mean
independent identically distributed Gaussian random tensor ${\bf Z}$,
or the sum of ${\bf Z}$ and a low rank deterministic tensor ${\bf X}_{0}$, i.e.
\begin{equation}
{\bf X}_{0}=\sum_{i=1}^{r}\lambda_{i}{\bf x}_{0}^{(1,i)}\otimes{\bf x}_{0}^{(2,i)}\otimes\ldots{\bf x}_{0}^{(d,i)}\label{eq:expre-X0}
\end{equation}
where $r$ is called the rank of ${\bf X}_{0}$. Here $(\lambda_{i})_{i=1,\ldots,r}$
are strictly positive real numbers, and for each $i=1,\ldots,r$ and
$k=1,...,d$, ${\bf x}_{0}^{(k,i)}$ is a $n_{i}\times1$ unit norm
vector. Recent works (see e.g \cite{hopkins,montanari-richard-2014,montanari-reichman-zeitouni-2017,perry-wein-bandeira}
) addressed teh detection/estimation of ${\bf X}_{0}$ when $r$ is reduced
to 1 and when the dimensions $n_{1},\ldots,n_{d}$ converge towards $\infty$
at the same rate. We also mention that \cite{hopkins} and \cite{perry-wein-bandeira}
only considered the case where the rank 1 tensor ${\bf X}_{0}$ is symmetric,
i.e. $n_{1}=n_{2}=\ldots=n_{d}$ and all vectors $({\bf x}_{0}^{(1,i)})_{i=1,\ldots,d}$
coincide. As the concept of singular value decomposition cannot be extended
to tensors, ad'hoc statistical strategies have been considered to  prove the (non)-existence of consistent tests:  \cite{montanari-richard-2014} and \cite{perry-wein-bandeira} established
that if $\lambda_1$ is larger than a certain upper bound, then  consistent detection of ${\bf X}_{0}$ is possible.
In the other direction, \cite{montanari-reichman-zeitouni-2017} and \cite{perry-wein-bandeira}
proved that if $\lambda_1$ is less than a certain lower bound (which is stricly less than the  above upper bound), then $\mathbf{X}_{0}$
is non-detectable in the sense that any test behaves as a random
choice between the two hypotheses. This is a remarkable phenomenon because such a
behaviour is not observed in the matrix case $d=2$. In effect, if
the largest eigenvalue of ${\bf X}_{0}$ is below $c^{1/4}$, then,
\cite{onatski-moreira-hallin-2013} proved when $r=1$ that
 there exist statistical tests having a better performance than 
a random choice, a result that \cite{montanari-reichman-zeitouni-2017} and \cite{perry-wein-bandeira}
obtained a different way in the symetric case. 

In \cite{hopkins}, \cite{montanari-richard-2014}, \cite{montanari-reichman-zeitouni-2017},
\cite{perry-wein-bandeira}  a main  assumption is 
that ${\bf X}_{0}$ is a rank 1 tensor. The purpose of the present
paper is to consider the case where $r\geq 1$: we  find out a sufficient condition on the parameters of ${\bf X}_{0}$ under which ${\bf X}_{0}$ is non-detectable. 
The problem of finding conditions under which the existence of a consistent detection detection is guaranteed 
is not addressed here.

\section{Model, notation, and background  } \label{sec:model}

The order-$d$ tensors are complex-valued, and it is assumed that $n_{1}=n_{2}=...=n$
in order to simplify the notations. 
The set $\varotimes^{d}\mathbb{C}^{n}$ is a complex vector-space
endowed with the standard scalar product 
\[
\forall\ \mathbf{X},\mathbf{Y}\in\varotimes^{d}\mathbb{C}^{n}\ \ \left\langle \mathbf{X},\mathbf{Y}\right\rangle =\sum_{i_{1},...,i_{d}}\mathbf{X}_{i_{1},...,i_{d}}\overline{\mathbf{Y}}_{i_{1},...,i_{d}}
\]
and the Frobenius norm $\left\Vert \mathbf{X}\right\Vert _{F}=\sqrt{\left\langle \mathbf{X},\mathbf{X}\right\rangle }.$ 

The spike (``the signal'') is assumed to be a tensor of fixed rank
$r$ following (\ref{eq:expre-X0}). Along this contribution, $n$ is large or, mathematically,
$n\to\infty$. We hence have for each $n$ a set of $n\times1$ vectors
$\left(\mathbf{x}_{0}^{(k,i)}\right)_{k=1...d, i=1,...,r}$. For each $k=1, \ldots, d$, 
we denote by $\boldsymbol{\chi}_{0}^{(k)}$ the $n \times r$ matrix 
$\boldsymbol{\chi}_{0}^{(k)} = (\mathbf{x}_{0}^{(k,1)}, \ldots, \mathbf{x}_{0}^{(k,r)})$. 
We impose
a non-erratic asymptotic behavior of the spike, and specifically, as all the
vectors $\mathbf{x}_{0}^{(k,i)}\in\mathbb{C}^{n\times1}$ have unit
norm, we suppose that for all $i,j$, $\left\langle \mathbf{x}_{0}^{(k,i)},\mathbf{x}_{0}^{(k,j)}\right\rangle = 
(\boldsymbol{\chi}_{0}^{(k)*}\boldsymbol{\chi}_{0}^{(k)})_{i,j}$
converges as $n\to\infty$. The rate of convergence is a technical
aspect that is out of the scope of this contribution: we will simply assume that the matrices
$(\boldsymbol{\chi}_{0}^{(k)*}\boldsymbol{\chi}_{0}^{(k)})_{k=1, \ldots, d}$ do not
depend on $n$. We define the SVD of $\boldsymbol{\chi}_{0}^{(k)}$
as $\mathbf{U}_{k}\left(\begin{array}{c}
\boldsymbol{\Sigma}_{k}\\
0
\end{array}\right)\mathbf{V}_{k}^{*}$ for $\mathbf{U}_{k}$ and $\mathbf{V}_{k}$ unitary matrices respectively
of size $n\times n$ and $r\times r$ and $\boldsymbol{\Sigma}_{k}$
a diagonal matrix with non-negative entries on the diagonal. $\mathbf{V}_{k}$
and $\boldsymbol{\Sigma}_{k}$ do no depend on $n$ because $\boldsymbol{\chi}_{0}^{(k)*}\boldsymbol{\chi}_{0}^{(k)} = {\bf V}_k \boldsymbol{\Sigma}_{k}^{2} {\bf V}_k^{*}$. 

We denote by $\mathbf{Z}$ the noise tensor, and assume that its entries are $\mathcal{N}_{\mathbb{C}}(0,1/n)$ independent identically distributed complex circular Gaussian random variables. 

In the following, we consider the alternative
$\mathcal{H}_{0}:\ \mathbf{Y}=\mathbf{Z}$ versus $\mathcal{H}_{1}:\ \mathbf{Y}=\mathbf{X}_{0}+\mathbf{Z}.$
We denote by  $p_{1,n}({\bf y})$ the probability probability density of $\mathbf{Y}$
under $\mathcal{H}_{1}$ and $p_{0,n}({\bf y})$ the density of $\mathbf{Y}$
under $\mathcal{H}_{0}$.  $\Lambda(\mathbf{Y})=\frac{p_{1}(\mathbf{Y})}{p_{0}(\mathbf{Y})}$ is 
the likelihood ratio and we denote by $\mathbb{E}_{0}$ the expectation under $\mathcal{H}_{0}$. 
 We now recall the fundamental information geometry results 
used in  \cite{montanari-reichman-zeitouni-2017} in order to address the detection problem.The following properties
are well known (see also \cite{banks-vershynin2017} section
3):
\begin{itemize}
\item (i) If $\mathbb{E}_{0}\left[\Lambda(\mathbf{Y})^{2}\right]$ is bounded, then no consistent detection test exists. 
\item (ii) If moroever $\mathbb{E}_{0}\left[\Lambda(\mathbf{Y})^{2}\right]=1+o(1)$, then the total variation distance between ${p}_{0,n}$ and ${p}_{1,n}$ converges towards $0$, and no test performs better than a decision at random. 
\end{itemize}
Therefore, the computation of the second order moment of $\Lambda(\mathbf{Y})$ under ${p}_{0,n}$ may provide 
 insights on the detection. We however notice 
that conditions (i) and (ii) are only sufficient. In particular, if $\limsup_n \mathbb{E}_{0}\left[\Lambda(\mathbf{Y})^{2}\right] = +\infty $, nothing can be inferred on the behaviour of the detection problem when 
$n \rightarrow +\infty$.

\section{Prior on the spike. Expression of the second-order moment. }

The density of $\mathbf{Z}$, seen as a collection of $n^{d}$ complex-valued
random variables, is obviously $p_{0,n}(\mathbf{z})=\kappa_{n}\exp\left(-n\left\Vert \mathbf{z}\right\Vert _{F}^{2}\right)$
where $\kappa_{n}=\left(\frac{n}{\pi}\right)^{n^{d}}$. On the one
hand, we notice that the second-order moment approach is not
suited to the deterministic model of the spike as presented previously.
Indeed, in this case $\mathbb{E}_{0}\left[\Lambda(\mathbf{Y})^{2}\right]$has
the simple expression $\exp\left(2n\left\Vert \mathbf{X}_{0}\right\Vert _{F}^{2}\right)$
and always diverges. 
On the other hand,
the noise tensor shows an invariance property: if $\boldsymbol{\Theta}_{1},...,\boldsymbol{\Theta}_{d}$
are unitary $n\times n$ matrices , then the density of the mode products
$\left(\boldsymbol{\Theta}_{1}\otimes\boldsymbol{\Theta}_{2}...\otimes\boldsymbol{\Theta}_{d}\right)\mathbf{Z}$
equals this of $\mathbf{Z}$. For $d=2$,the notation  $\left(\boldsymbol{\Theta}_{1}\otimes\boldsymbol{\Theta}_{2}\right)\mathbf{Z}$ simply means  $\boldsymbol{\Theta}_{1}\mathbf{Z}\boldsymbol{\Theta}_{2}$
and for a general $d$, $\left(\left(\boldsymbol{\Theta}_{1}\otimes\boldsymbol{\Theta}_{2}...\otimes\boldsymbol{\Theta}_{d}\right)\mathbf{Z}\right)_{i_{1},...,i_{d}}$
is
\[
\sum_{\ell_{1},...,\ell_{d}}\left(\boldsymbol{\Theta}_{1}\right)_{i_{1},\ell_{1}}\left(\boldsymbol{\Theta}_{2}\right)_{i_{2},\ell_{2}}...\left(\boldsymbol{\Theta}_{d}\right)_{i_{d},\ell_{d}}\mathbf{Z}_{\ell_{1},...,\ell_{d}}.
\]
We hence  modify the data according to the procedure: we
pick i.i.d. complex Haar samples $\boldsymbol{\Theta}_{1},...,\boldsymbol{\Theta}_{d}$
and change the data tensor $\mathbf{Y}$ into 	 $\left(\boldsymbol{\Theta}_{1}\otimes\boldsymbol{\Theta}_{2}...\otimes\boldsymbol{\Theta}_{d}\right)\mathbf{Y}.$
This does not affect the distribution of the noise, but this amounts
to assume a prior on the spike. Indeed, the vectors $\mathbf{x}_{0}^{(k,i)}$
are replaced by $\boldsymbol{\Theta}_{k}\mathbf{x}_{0}^{(k,i)}$.
They are all uniformly distributed on the unit sphere of $\mathbb{C}^{n}$
and for $ k \neq l$, vectors $\boldsymbol{\Theta}_{k}\mathbf{x}_{0}^{(k,i)}$
and $\boldsymbol{\Theta}_{l}\mathbf{x}_{0}^{(l,j)}$ are independent for each $i,j$. However, 
vectors $(\boldsymbol{\Theta}_{k}\mathbf{x}_{0}^{(k,i)})_{i=1, \ldots, r}$ are not independent. 
In the following, the data
and the noise tensors after this procedure are still denoted respectively
by $\mathbf{Y}$ and $\mathbf{Z}$. 

We are now in position to give a closed-form expression of the second-order
moment of ${\boldsymbol \Lambda}({\bf Y})$ . 
We have $p_{1,n}(\mathbf{Y})=\mathbb{E}_{X}\left[p_{0,n}(\mathbf{Y}-\mathbf{X})\right]$
where $\mathbb{E}_X$ is the mathematical expectation over the distribution of the spike, 
or equivalently over the Haar matrices $(\boldsymbol{\Theta}_{k})_{k=1, \ldots, d}$. It holds that 
\begin{multline*}
\mathbb{E}_{0}\left[\Lambda(\mathbf{Y})^{2}\right]  =  \mathbb{E}_{X,X'}\left[\exp\left(2n\mathfrak{R}\left\langle \mathbf{X},\mathbf{X}'\right\rangle \right)\right] = \\
\mathbb{E}_{X,X'}\left[\exp\left(2n\mathfrak{R}\sum_{i,j=1}^{r}\lambda_{i}\lambda_{j}\prod_{k=1}^{d}\left\langle \left(\mathbf{\Theta}'_{k}\right)^{*}\mathbf{\Theta}_{k}\mathbf{x}_{0}^{(k,i)},\mathbf{x}_{0}^{(k,j)}\right\rangle \right)\right]
\end{multline*}
where $\mathbb{E}_{X,X'}$ is over independent
copies $\mathbf{X},\mathbf{X}'$ of the spike associated respectively
with $(\boldsymbol{\Theta}_{k})_{k=1, \ldots, d}$ and $(\boldsymbol{\Theta}_{k}')_{k=1, \ldots, d}$. $\mathfrak{R}$ stands for the real part. As $\mathbf{\Theta}_{k}$ and $\mathbf{\Theta}_{k}'$ are Haar and
independent, then $\left(\mathbf{\Theta}'_{k}\right)^{*}\mathbf{\Theta}_{k}$
is also Haar distributed and 
$\mathbb{E}_{0}\left[\Lambda(\mathbf{Y})^{2}\right]  =\mathbb{E}\left[\exp\left(2n\eta\right)\right]$,
where the expectation is over the i.i.d. Haar matrices $\mathbf{\Theta}_{1},\mathbf{\Theta}_{2},...,\mathbf{\Theta}_{d}$
and 
\begin{equation}
\eta=\mathfrak{R}\sum_{i,j=1}^{r}\lambda_{i}\lambda_{j}\prod_{k=1}^{d}\underbrace{\left\langle \mathbf{\Theta}_{k}\mathbf{x}_{0}^{(k,i)},\mathbf{x}_{0}^{(k,j)}\right\rangle }_{\xi_{k}^{(i,j)}}.\label{eq:eta expanded}
\end{equation}
$\eta$ may be factored as
$\eta=\mathfrak{R}\left[\boldsymbol{\lambda}^{T}\left(\varodot_{k=1}^{d}\left(\boldsymbol{\chi}_{0}^{(k)*}\mathbf{\boldsymbol{\Theta}}_{k}\boldsymbol{\chi}_{0}^{(k)}\right)\right)\mathbf{\boldsymbol{\lambda}}\right].$
In the latter equation, $\varodot$ stands for the Hadamard product
of matrices. The ultimate simplification comes from the SVD of $\boldsymbol{\chi}_{0}^{(k)}$:
\[
\boldsymbol{\chi}_{0}^{(k)*}\mathbf{\boldsymbol{\Theta}}_{k}\boldsymbol{\chi}_{0}^{(k)}=\mathbf{V}_{k}\left(\begin{array}{cc}
\mathbf{\boldsymbol{\Sigma}}_{k} & \mathbf{0}\end{array}\right)\mathbf{U}_{k}^{*}\mathbf{\boldsymbol{\Theta}}_{k}\mathbf{U}_{k}\left(\begin{array}{c}
\mathbf{\boldsymbol{\Sigma}}_{k}\\
\mathbf{0}
\end{array}\right)\mathbf{\boldsymbol{\Sigma}}_{k}\mathbf{V}_{k}^{*}.
\]
 Firstly, $\mathbf{U}_{k}^{*}\mathbf{\boldsymbol{\Theta}}_{k}\mathbf{U}_{k}$
has the same distribution as $\boldsymbol{\Theta}_{k}$; secondly,
we may associate with any $\boldsymbol{\Theta}_{k}$ its upper $r\times r$ block,
that we will denote $\boldsymbol{\Psi}_{k}.$ As a conclusion, we
may express $\eta$ as 
\begin{equation}
\eta=\mathfrak{R}\left[\boldsymbol{\lambda}^{T}\left(\varodot_{k=1}^{d}\left(\mathbf{V}_{k}\mathbf{\boldsymbol{\Sigma}}_{k}\boldsymbol{\Psi}_{k}\mathbf{\boldsymbol{\Sigma}}\mathbf{V}_{k}^{*}\right)\right)\mathbf{\boldsymbol{\lambda}}\right].\label{def:eta}
\end{equation}

\section{Extending known results}

When $r=1$, Montanari et al. \cite{montanari-reichman-zeitouni-2017}
found a bound on the parameter $\lambda_1$ ensuring that $\mathbb{E}_{0}\left[\Lambda(\mathbf{Y})^{2}\right]$
is bounded. In this simple case,  $\eta$ has a simple expression since 
 $\eta=\lambda^{2}\mathfrak{R}\prod_{k=1}^{d}\xi_{k}$ where
the $(\xi_{k})_{k=1, \ldots, d}$  are i.i.d. distributed as the first component of a
uniform vector of the unit sphere of $\mathbb{C}^{n}$. As in \cite{montanari-reichman-zeitouni-2017},
we introduce 
\begin{align}
\beta_{d}^{\text{2nd}} & =\sqrt{\min_{u\in[0,1]}-\frac{1}{u^{d}}\log(1-u^{2})}.\label{eq:beta d}
\end{align}
 Adapting the result of the mentionned article the complex-circular
context is straight-forward:
\begin{thm}[case r=1 (Montanari et al.)]
\label{theo:montanari}
 Let $\xi_{1},...,\xi_{d}$ be i.i.d. distributed as the first component
of a vector uniformly distributed on the unit sphere of $\mathbb{C}^{n}$. 

If $\text{\ensuremath{\lambda_1}}<\sqrt{\frac{d}{2}}\beta_{d}^{\text{2nd}}$
then $\mathbb{E}_{0}\left[\exp\left(2n\lambda_1^{2}\mathfrak{R}\prod_{k=1}^{d}\xi_{k}\right)\right]$
is bounded; moreover, if $d>2$, the above expectation is 1+o(1).\label{thm: montanari}
\end{thm}

This non-obvious result may be used in order to derive a condition ensuring that hypotheses $\mathcal{H}_0$ 
and  $\mathcal{H}_1$ are indistinguishable when $r>1.$ In this respect, recall the expansion (\ref{eq:eta expanded}).
Thanks to  the H\"{o}lder inequality, $\mathbb{E}_{0}\left[\Lambda(\mathbf{Y})^{2}\right]$
is upper bounded by (see \eqref{eq:eta expanded} for the definition of $\xi_k^{(i,j)}$)
\begin{equation}
\label{eq:holder}
\prod_{i,j=1}^{r}\mathbb{E}^{1/p_{i,j}}\left[\exp\left(2np_{i,j}\lambda_{i}\lambda_{j}\mathfrak{R}\prod_{k=1}^{d}\xi_k^{(i,j)}\right)\right]
\end{equation}
for any non-negative numbers $p_{i,j}$ such that $\sum_{i,j}\frac{1}{p_{i,j}}=1.$
For fixed $i,j$, we notice that  the random variables $(\xi_k^{(i,j)})_{k=1, \ldots,d}$ 
verify the condition of Theorem \ref{thm: montanari}. Any of the
expectations in (\ref{eq:holder}) are hence upper-bounded when $n\to\infty$
provided that, for all $i,j$: $p_{i,j}\lambda_{i}\lambda_{j}<\frac{d}{2}\left(\beta_{d}^{\text{2nd}}\right)^{2}$.
Choosing eventually $p_{i,j}=\frac{\left(\sum_{p}\lambda_{p}\right)^{2}}{\lambda_{i}\lambda_{j}}$,
we deduce 
\begin{thm}[case $r\geq1$ - extension of Theorem \ref{thm: montanari} ]
\label{th:suboptimum}
If $\sum_{i=1}^{r} \lambda_{i}<\sqrt{\frac{d}{2}}\beta_{d}^{\text{2nd}}$
then $\mathbb{E}_{0}\left[\Lambda(\mathbf{Y})^{2}\right]$ is bounded.
If moreover $d>2$, we have $\mathbb{E}_{0}\left[\Lambda(\mathbf{Y})^{2}\right]=1+o(1)$\label{thm:Hoelder} and the hypotheses $\mathcal{H}_0$ and $\mathcal{H}_1$ are indistinguishable. 
\end{thm}

\begin{rem}
	Due to the use of the H\"{o}lder inequality, Theorem \ref{th:suboptimum} is  suboptimum in general. 
	The inequality is patently an equality when 
	$\forall k,i,j$, ${\bf x}_0^{(k,i)}={\bf x}_0^{(k,j)}$, i.e. the spike has rank $r=1$ and amplitude $\sum_{i=1}^{r}\lambda_i$. 
\end{rem}

\section{A tighter bound}

The main result of our contribution is the following
\begin{thm}[case $r\geq1$]
We define $\eta_{\max}$ as 
\begin{equation}
\label{eq:def-etamax}
\eta_{\max}=\text{\textbf{\ensuremath{\boldsymbol{\lambda}}}}\left(\varodot_{k=1}^{d}\mathbf{\boldsymbol{\chi}}_{0}^{(k)*}\mathbf{\boldsymbol{\chi}}_{0}^{(k)}\right)\mathbf{\boldsymbol{\lambda}.}
\end{equation}
If $\sqrt{\eta_{\max}}<\sqrt{\frac{d}{2}}\beta_{d}^{\text{2nd}}$
then, for  $d>2$, $\mathbb{E}_{0}\left[\Lambda(\mathbf{Y})^{2}\right]=1+o(1).$
\label{thm:main}
\end{thm}

Before providing elements of the proof of the above result, we may
briefly justify why the bound in Theorem \ref{thm:main} is tighter
than this of Theorem \ref{thm:Hoelder}, whatever the choice of $\boldsymbol{\lambda}.$
On the one hand, indeed, $\left(\sum_{i}\lambda_{i}\right)^{2}=\boldsymbol{\lambda}^{T}\mathbf{J}\boldsymbol{\lambda}$
where $\mathbf{J}$ is the $r\times r$ matrix having all its entries
equal to $1$. On the other hand, all the vectors $\mathbf{x}_{0}^{(k,i)}$
are normalized and consequently, any of the diagonal entries of $\mathbf{\boldsymbol{\chi}}_{0}^{(k)*}\mathbf{\boldsymbol{\chi}}_{0}^{(k)}$
equals $1$ and for any $i\neq j$, $\left|(\mathbf{\boldsymbol{\chi}}_{0}^{(k)*}\mathbf{\boldsymbol{\chi}}_{0}^{(k)})_{i,j}\right|\leq1.$
This proves that $\left(\sum_{i}\lambda_{i}\right)^{2}-\eta_{\max}=\boldsymbol{\lambda}^{T}\left(\mathbf{J}-\varodot_{k=1}^{d}\mathbf{\boldsymbol{\chi}}_{0}^{(k)*}\mathbf{\boldsymbol{\chi}}_{0}^{(k)}\right)\boldsymbol{\lambda}\geq0.$ 

We provide the key elements of the proof of Theorem \ref{thm:main}.
Remind that we are looking for a condition on the spike under which
$\mathbb{E}\left[\exp\left(2n\eta\right)\right]$ is bounded. Evidently,
the divergence may occur only when $\eta>0.$ We hence consider $E_{1}=\mathbb{E}\left[\exp\left(2n\eta\right)\mathbf{1}_{\eta>\epsilon}\right]$
and $E_{2}=\mathbb{E}\left[\exp\left(2n\eta\right)\mathbf{1}_{\eta\leq\epsilon}\right]$, and prove that under the condition $\sqrt{\eta_{\max}}<\sqrt{\frac{d}{2}}\beta_{d}^{\text{2nd}}$,  for a certain small enough $\epsilon$, 
 $E_1 = o(1)$  (for $d\geq 2$) and that $E_2 = 1 + o(1)$ (for $d>2$).

\paragraph*{The $E_{1}$ term. }

It is clear that the boundedness of the integral $E_{1}$ is achieved
when $\eta$ rarely deviates from $0$. As remarked in \cite{montanari-reichman-zeitouni-2017}, the natural machinery to consider
to understand $E_{1}$ is this of the Large Deviation Principle (LDP). In
essence, if $\eta$ follows the LDP with rate $n$, there can be found
a certain non-negative function called Good Rate Function (GRF) $I_{\eta}$
such that for any Borel set $A$ of $\mathbb{R}$, $\frac{1}{n}\log\mathbb{P}\left(\eta\in A\right)$
converges towards $\sup_{x\in A}-I_{\eta}(x)$. The existence of a GRF allows one to analyze the asymptotic 
behaviour of integral $E_1$. Indeed, the Varadhan lemma (see Theorem 4.3.1 in \cite{dembo-zeitouni2009}) states that $\frac{1}{n}\log\mathbb{E}\left[\exp\left(2n\eta\right)\mathbf{1}_{\eta>\epsilon}\right]\to\sup_{x>\epsilon}\left(2x-I_{\eta}(x)\right)$
and hence the $E_{1}$ term converges towards $0$ when $\sup_{x>\epsilon}\left(2x-I_{\eta}(x)\right) < 0$.

We thus justify that $\eta$ follows a Large Deviation Principle with rate $n$, and we compute a lower bound of its GRF. For this, we use that for each $k$, random matrix $\boldsymbol{\Psi}_{k}$ defined in (\ref{def:eta})
follows a LDP with rate $n$ and that its GRF at the parameter $\boldsymbol{\psi}\in\mathbb{C}^{r\times r}$ (we may evidently take $\left\Vert \boldsymbol{\psi}\right\Vert _{2}\leq1$)
is $\log\det\left(\mathbf{I}_{r}-\boldsymbol{\psi}^{*}\boldsymbol{\psi}\right)$ (see Theorem 3-6 in \cite{gamboa-2014}). $\eta$ is a function of the i.i.d. matrices $(\boldsymbol{\Psi}_{k})_{k=1, \ldots, d}$. Therefore, the contraction
principle (see Theorem 4.2.1 in \cite{dembo-zeitouni2009}) ensures 
that $\eta$ follows a LDP with rate $n$ and GRF $I_{\eta}$ given, for each real $x$ 
in the range of $\eta$, as the  solution of the optimization problem:
\begin{align}
 & \max_{\forall k\ 0\leq\alpha_{k}\leq1}\max_{\begin{array}{c}
\forall k\left\Vert \boldsymbol{\psi}_{k}\right\Vert =\alpha_{k}\\
\eta(\boldsymbol{\psi}_{1},...,\boldsymbol{\psi}_{d})=x
\end{array}}\ \ \sum_{k=1}^{d}\log\det\left(\mathbf{I}_{r}-\boldsymbol{\psi}_{k}^{*}\boldsymbol{\psi}_{k}\right).\label{eq:optim}
\end{align}

When $d \geq 3$, the solution of this optimization problem cannot apparently be expressed in closed form. We thus just provide a lower bound of $I_{\eta}(x)$. When $d=2$, it is possible to evaluate $I_{\eta}(x)$, but due to the lack of space, we do not report 
the corresponding result in the present paper. 
\begin{prop} 
\label{prop:upper-bound-Ieta}
For each $x \in \mathbb{R}$, it holds that 
\begin{equation}
\label{eq:upperbound-Ieta}
I_{\eta}(x)\geq - d\log\left(1-\left(\frac{\left|x\right|}{\eta_{\max}}\right)^{2/d}\right).
\end{equation}
where the right-hand side should be understood as $+\infty$ if $|x| \geq \eta_{\max}$. 
\end{prop}
In order to establish Proposition \ref{prop:upper-bound-Ieta}, we use the following algebraic result whose proof is omitted. 
\begin{lem}
\label{lem:magie noire}For any matrices $(\mathbf{A}_{k})_{k=1, \ldots, d} \in\mathbb{C}^{r\times r}$
and vector $\boldsymbol{\mathbf{\lambda}} \in \mathbb{R}^{r}$, the supremum of $\left|\mathbf{\boldsymbol{\lambda}}^{T}\varodot_{k=1}^{d}\left(\mathbf{A}_{k}\boldsymbol{\psi}_{k}\mathbf{A}_{k}^{*}\right)\mathbf{\boldsymbol{\lambda}}\right|$
over $r\times r$ matrices $\boldsymbol{\psi}_{k}$ such that for
all k: $||\boldsymbol{\psi}_{k}||_{2}=\alpha_{k}$ is 
\[
\left(\prod_{k=1}^{d}\alpha_{k}\right)\mathbf{\boldsymbol{\lambda}}^{T}\left(\varodot_{k=1}^{d}\left(\mathbf{A}_{k}\mathbf{A}_{k}^{*}\right)\right)\mathbf{\boldsymbol{\lambda}}.
\]
\end{lem}
The immediate consequence of this lemma is that the random variable
$\eta$ is bounded and $\left|\eta\right|\leq\eta_{\max}$ where $\eta_{\max}$
is given by (\ref{eq:def-etamax}). Moreover, take a set of matrices
$\boldsymbol{\psi}_{k}$ such that $\left\Vert \boldsymbol{\psi}_{k}\right\Vert_2	 =\alpha_{k}\in[0,1];$
then by Lemma \ref{lem:magie noire}, $\left|\eta\left(\boldsymbol{\psi}_{1},...,\boldsymbol{\psi}_{d}\right) \right|\leq\left(\prod_{k}\alpha_{k}\right)\eta_{\max}$
hence the optimization (\ref{eq:optim}) is to be carried out only
on the set of matrice $\boldsymbol{\psi}_{k}$ such that $\prod_{k}\alpha_{k}\geq\frac{\left|x\right|}{\eta_{\max}}.$
On the other hand, one may use the generous bound $\log\det\left(\mathbf{I}_{r}-\boldsymbol{\psi}_{k}^{*}\boldsymbol{\psi}_{k}\right)\leq\log\left(1-\left\Vert \boldsymbol{\psi}_{k}\right\Vert _{2}^{2}\right)$
and finally prove that 
\[
-I_{\eta}(x)\leq\max_{\prod_{k}\alpha_{k}\geq\frac{\left|x\right|}{\eta_{\max}}}\sum_{k=1}^{d}\log\left(1-\alpha_{k}^{2}\right).
\]
The supremum of the r.h.s. of this equation is achieved for balanced
$\alpha_{k}$ and we immediately obtain (\ref{eq:upperbound-Ieta}). This completes the proof of 
Proposition \ref{prop:upper-bound-Ieta}. 
\paragraph*{The $E_{1}$ term.}
We are now in position to conclude that $E_1 = o(1)$. Varadhan's lemma implies that
$\frac{1}{n}\log E_{1}\to\sup_{x\geq\epsilon}\left[2x-I_{\eta}(x)\right]$. Using 
Proposition \ref{prop:upper-bound-Ieta} and setting $u=\left(\frac{\left|x\right|}{\eta_{\max}}\right)^{1/d}$, we obtain immediately that for each $\delta > 0$,  $\frac{1}{n}\log E_{1}$ is less than
\[
\frac{1}{n}\log E_{1} < \sup_{u\geq\tilde{\epsilon}}\left[2u^{d}\left(\eta_{\max}+\frac{d}{2}\frac{1}{u^{d}}\log(1-u^{2})\right)\right] + \delta
\]
for $n$ large enough, where $\tilde{\epsilon}=\left(\epsilon/\eta_{\max}\right)^{1/d}$. Recalling (\ref{eq:beta d})
and choosing $\delta$ small enough, 
we deduce that the condition $\eta_{\max}<\frac{d}{2}\left(\beta_{d}^{\text{2nd}}\right)^{2}$
implies that $E_{1}\to0$. This holds for any order $d\geq2$. 

\paragraph*{The $E_{2}$ term.}

The Varadhan lemma may be invoked: but its conclusion, namely $\frac{1}{n}\log E_{2}\to0$,
says nothing on the boundedness of $E_{2}.$ We have, however

\begin{eqnarray*}
E_{2} & = & \int_{0}^{\infty}\mathbb{P}\left(\exp(2n\eta)\geq t\ \text{ and }\eta\leq\epsilon\right)\text{d}t\\
 & = & \int_{-\infty}^{0}\mathbb{P}\left(\eta\geq u\ \text{ and }\eta\leq\epsilon\right)2n\exp(2nu)\text{d}u+\\
 &  & \int_{0}^{\epsilon}\mathbb{P}\left(\eta\geq u\ \text{ and }\eta\leq\epsilon\right)2n\exp(2nu)\text{d}u\\
 & \leq & \mathbb{P}\left(\eta\leq\epsilon\right)+\int_{0}^{\epsilon}\mathbb{P}\left(\eta\geq u\right)2n\exp(2nu)\text{d}u.
\end{eqnarray*}
A weak consequence of the LDP on $\eta$ is the concentration of $\eta$
around $0$, namely $\mathbb{P}(\eta\leq\epsilon)=1-\mathbb{P}(\eta>\epsilon)=1-o(1).$
We recall the expanded expression for $\eta$: see (\ref{eq:eta expanded}).
Notice that $\eta\geq u$ implies that at least one of the $r^{2}$
terms of this expansion is at least equal to $\frac{u}{r^{2}}.$ By
the union bound, and the fact that $\mathfrak{R}\prod_{k=1}^{d}\xi_{k}^{(i,j)}\leq\prod_{k=1}^{d}\left|\xi_{k}^{(i,j)}\right|$
we deduce that $\mathbb{P}\left(\eta\geq u\right)\leq\sum_{i,j=1}^{r}\mathbb{P}\left(\prod_{k=1}^{d}\left|\xi_{k}^{(i,j)}\right|\geq\frac{u}{r^{2}\lambda_{i}\lambda_{j}}\right).$
Invoking again the union bound and noticing that for fixed $i,j$, $\left( \xi_{k}^{(i,j)}\right)_{k=1,...,d}$ have the same distribution,  we deduce that 
\[
\mathbb{P}\left(\eta\geq u\right)\leq d\sum_{i,j=1}^{r}\mathbb{P}\left(\left|\xi_{k}^{(i,j)}\right|\geq\left(\frac{u}{r^{2}\lambda_{i}\lambda_{j}}\right)^{1/d}\right).
\]
Now, the density of $\xi_{k}^{(i,j)}$ is in polar coordinates $\frac{n-1}{\pi}\left(1-r^{2}\right)^{n-2}$
hence, choosing $\epsilon$ such that $\epsilon \leq r^2 \max_{i,j}\lambda_i \lambda_j$: 

 $\mathbb{P}\left(\left|\xi_{k}^{(i,j)}\right|\geq\left(\frac{u}{r^{2}\lambda_{i}\lambda_{j}}\right)^{1/d}\right)=\left(1-\left(\frac{u}{r^{2}\lambda_{i}\lambda_{j}}\right)^{2/d}\right)^{n-1}.$
For any $0\leq x<1$, $\log(1-x)\leq-x$, hence
\begin{align*}
E_{2} & \leq d\sum_{i,j}2n\int_{0}^{\epsilon}\exp\left(-(n-1)\left(\frac{u}{r^{2}\lambda_{i}\lambda_{j}}\right)^{2/d}+2nu\right)du.
\end{align*}
When $d>2$, it is always possible to determine $\epsilon$ sufficiently
small such that $-(n-1)\left(\frac{u}{r^{2}\lambda_{i}\lambda_{j}}\right)^{2/d}+2nu\leq-\frac{n-1}{2}\left(\frac{u}{r^{2}\lambda_{i}\lambda_{j}}\right){}^{2/d}.$
This implies that, for such an $\epsilon,$ we have 
\[
E_{2}\leq
d^{2}r^{2}
n\left(\frac{2}{n-1}\right)^{d/2}
\sum_{i,j}\lambda_{i}\lambda_{j}
\int_{0}^{\infty}v^{d/2-1}\exp(-v)dv.
\]
The r.h.s. is of course $o(1)$ since $d>2$.

\begin{rem}
\label{re:not-tight}
The bound $\sqrt{\eta_{\max}}<\sqrt{\frac{d}{2}}\beta_{d}^{\text{2nd}}$
 guarantees the non-detectability but it is not tight in general because, in order to study the asymptotics of $E_1$, we replaced
 the true GRF $I_{\eta}$ by the lower bound (\ref{eq:upperbound-Ieta}).
Based on the loose inequality $\log\det\left(\mathbf{I}_{r}-\boldsymbol{\psi}_{k}^{*}\boldsymbol{\psi}_{k} \right)\leq\log\det\left(1-\left\Vert \boldsymbol{\psi}_{k}\right\Vert _{2}^{2}\right)$,  (\ref{eq:upperbound-Ieta}) may not be very accurate. It is easy to check that the equality is reached in (\ref{eq:upperbound-Ieta})
when all the matrices $(\boldsymbol{\chi}_{0}^{(k)})_{k=1, \ldots, d}$ are rank 1, i.e. if the rank of ${\bf X}_0$ is equal to 1. Therefore, 
the lower bound  (\ref{eq:upperbound-Ieta}) of $I_{\eta}$ is all the better as all the matrices
$\boldsymbol{\chi}_{0}^{(k)}$ are close to being rank 1 matrices. This suggests that, conversely, the bound  (\ref{eq:upperbound-Ieta}) is likely to be loose when matrices $(\boldsymbol{\chi}_{0}^{(k)})_{k=1, \ldots, d}$ are close to be orthogonal. As an illustration, we would like to consider experimental results. For a given configuration of the spike, we have chosen at random the matrices $\boldsymbol{\psi}_{k}$ with $\| \boldsymbol{\psi}_{k} \| \leq  1$. For each trial, we plot the points of coordinates $x= \eta(\boldsymbol{\psi}_1,...,\boldsymbol{\psi}_d)$
 and $y=\sum_{k=1}^{d}\log\det\left(\mathbf{I}_{r}-\boldsymbol{\psi}_{k}^{*}\boldsymbol{\psi}_{k}\right)$
and we obtain a cloud the upper envelope of which is a representation of the true GRF of $\eta$; for comparison, we have plotted the graph of the function defined by the lower bound (\ref{eq:upperbound-Ieta}). We have chosen $r=2$, $d=3$,  and two configurations of the spike: in the first one, all the matrices $\boldsymbol{\chi}_k$ have orthogonal columns (top graph of \ref{fig:res}), in the second one, the eigenvalues of $\boldsymbol{\chi}_k^*\boldsymbol{\chi}_k$ are the same for $k=1,2$ equal to $1.8$ and $0.2$ (bottom graph of \ref{fig:res}). 
\begin{figure}[htb]
	\begin{minipage}[b]{1\linewidth}
		\centering
		\centerline{\includegraphics[scale=0.25]{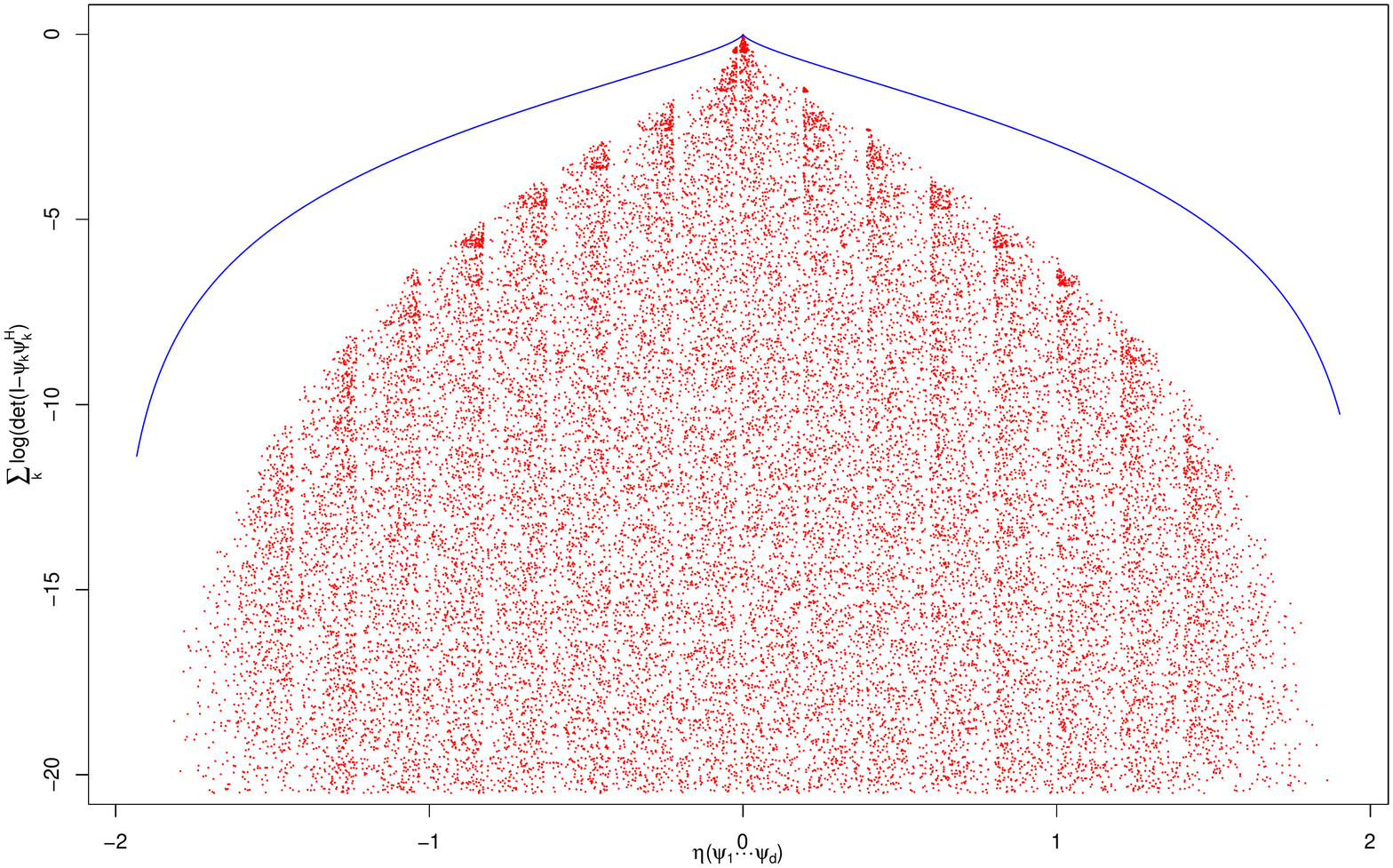}}
		\centering
		\end{minipage}	
		\begin{minipage}[b]{1\linewidth}
		\centerline{\includegraphics[scale=0.25]{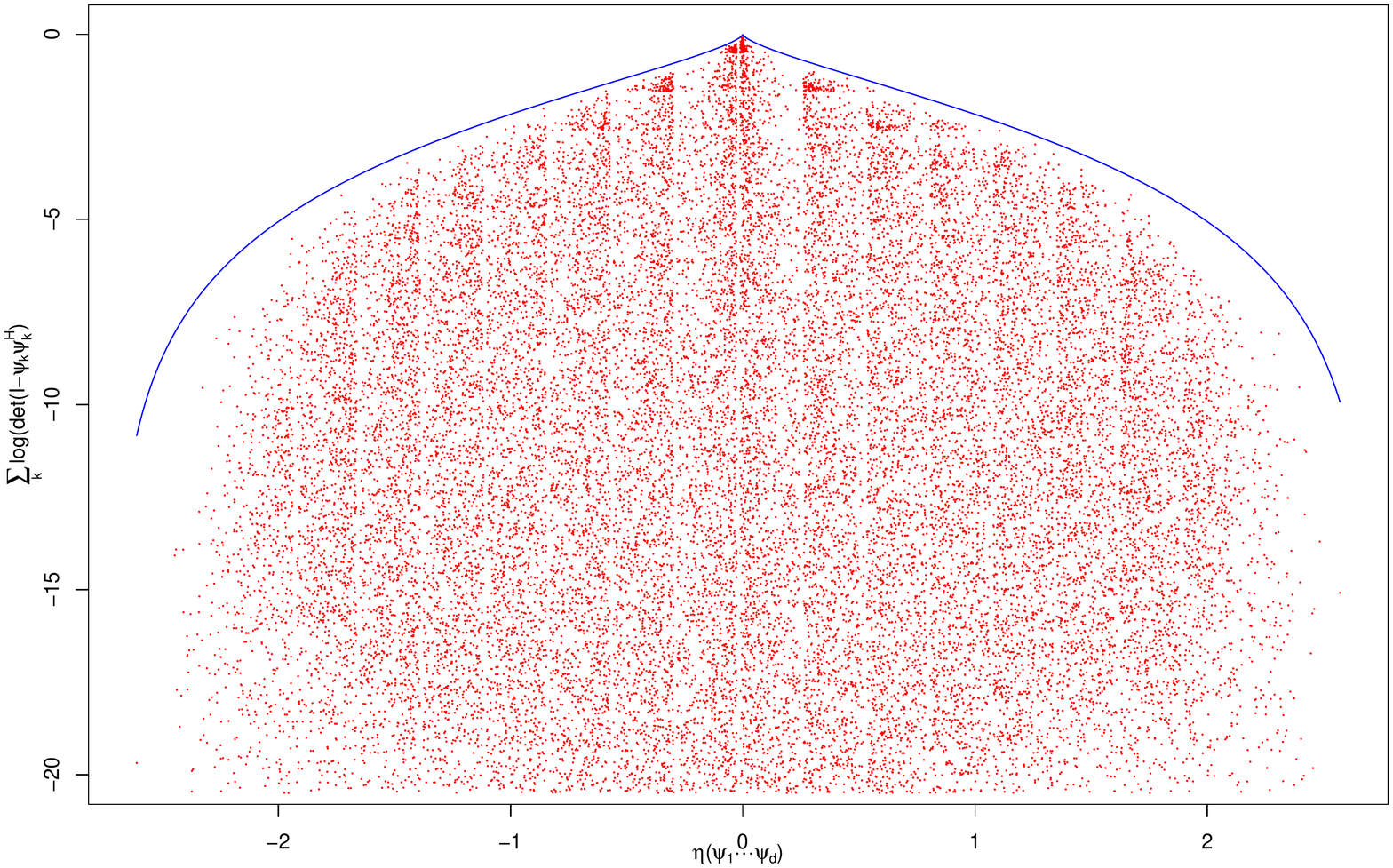}}
	\end{minipage}
	\caption{ $-I_{\eta }$ and our upper bound }
	\label{fig:res}
\end{figure}
\end{rem}
\begin{rem}
\label{re:d=2}
In the specific case $d=2$, it is possible to compute in closed-form the exact GRF $I_{\eta}$ of $\eta$, and to establish the following result:  if 
 $\mu_{\max}({\bf X}_0 {\bf X}_0^*)<\beta_2^{\text{2nd}} =1 $ (here, $\mu_{\max }$ denotes the largest eigenvalue), then $E_1$ converges
towards $0$. The approach we used in this paper to upper-bound $E_2$ for $d > 2$ is unsuccesfull for $d=2$. However, it is possible to adapt the technique used in \cite{montanari-reichman-zeitouni-2017}:  if $\mu_{\max}({\bf X}_0 {\bf X}_0^*)<1 $, then $E_2$ is bounded. From both results,  it may be concluded 
 that under the condition $\mu_{max}({\bf X}_0 {\bf X}_0^*) < 1$,  no consistent detection test can be found. 
\end{rem}

\section{Conclusion}
In this paper, we have addressed the detection problem of a rank $r$ high-dimensional tensor ${\bf X}_0$. We have generalized the results of 
\cite{montanari-reichman-zeitouni-2017} to the case where $r>1$, and established that if parameter $\eta_{max}$ defined by 
(\ref{eq:def-etamax}) is less that parameter $\beta_2^{\text{2nd}}$ introduced in \cite{montanari-reichman-zeitouni-2017}, 
the low rank tensor is undetectable. This condition is based on the lower bound (\ref{eq:upperbound-Ieta}) of the GRF $I_{\eta}$ which is however not tight in general. It is thus relevant to try to improve this bound in a future work.

\bibliographystyle{ieeetr}
\bibliography{biblioRank1}

\end{document}